\documentclass[printer]{aa} 
\usepackage{graphicx}
\usepackage{txfonts,natbib}

\def\gx{\mbox{GX 339-4}}

\def\hdsqt{\mbox{H\,1743-322}}

\def\xtejqcq{\mbox{XTE J1550-564}}
\def\xtejscq{\mbox{XTE J1650-500}}

\def\cmmoinsdeux{\mbox{ cm}^{-2}}

\def\microns{\mbox{ } \mu \mbox{m}}

\def\mags{\mbox{ magnitudes}}

\def\hour{^{h}}

\def\min{^{m}}

\def\secp{{\rlap.}^{s}}

\def\adeg{^{\circ}}

\def\amin{^\prime}
\def\aminp{{\rlap.}^{\prime}}
\def\asec{^{\prime \prime}}
\def\asecp{{\rlap.}^{\prime \prime}}

\def\Av{A_{\rm V}}
\def\Aj{A_{\rm J}}
\def\Ah{A_{\rm H}}
\def\Ak{A_{\rm K}}

\def\nh{N_{\rm H}}

\def\ltsima{\; \buildrel < \over \sim \;}
\def\simlt{\lower.5ex\hbox{\ltsima}}            
\def\gtsima{\; \buildrel > \over \sim \;}
\def\simgt{\lower.5ex\hbox{\gtsima}}            

\begin{document}
\title{Infrared study of $\hdsqt$ in outburst: a radio-quiet and NIR-dim microquasar\thanks{Based on observations collected at the European Southern Observatory, Chile, through programs 071.D-0073, 073.D-0341 and 081.D-0401.}}

\author{S. Chaty\inst{1,2} \and A.J. Mu\~noz Arjonilla\inst{3} \and G. Dubus\inst{4,5}}

\institute{AIM (UMR-E 9005 CEA/DSM-CNRS-Universit\'e Paris Diderot),
Irfu/Service d'Astrophysique, Centre de Saclay,
FR-91191 Gif-sur-Yvette Cedex, France, \email{chaty@cea.fr}
\and
Institut Universitaire de France, 103, boulevard Saint-Michel 75005 Paris, France
\and Grupo de Investigaci\'on FQM-322, Universidad de Ja\'en, Campus Las Lagunillas s/n A3-065, E-23071 - Ja\'en, Spain, \email{ajmunoz@ujaen.es}
\and Univ. Grenoble Alpes, IPAG, F-38000 Grenoble, France, \email{guillaume.dubus@obs.ujf-grenoble.fr} 
\and CNRS, IPAG, F-38000 Grenoble, France
}

\date{Received March 15, 2013; accepted April 15, 2013}



\abstract
{{Microquasars are accreting Galactic sources that are commonly observed to launch relativistic jets. One of the most important issues regarding these sources is the energy budget of ejections relative to the accretion of matter.}}
{The X-ray binary, black hole candidate, and microquasar $\hdsqt$ exhibited a series of X-ray outbursts between 2003 and 2008. We took optical and near-infrared (OIR) observations with the ESO/NTT telescope during three of these outbursts (2003, 2004, and 2008). The goals of these observations were to investigate the presence of a jet, and to disentangle the various contributions constituting the spectral energy distribution (SED): accretion, ejection, and stellar emission.}
{Photometric and spectroscopic OIR observations allowed us to produce a high time-resolution lightcurve in K$_s$-band, to analyze emission lines present in the IR spectra, to construct a multiwavelength SED including radio, IR, and X-ray data, and to complete the OIR vs X-ray correlation of black hole binaries with $\hdsqt$ data points.}
{We detect rapid flares of duration $\sim 5$\,minutes in the high time-resolution IR lightcurve. We identify hydrogen and helium emission lines in the IR spectra, coming from the accretion disk. The IR SED exhibits the spectral index typically associated with the X-ray high, soft state in our observations taken during the 2003 and 2004 outbursts, while the index changes to one that is typical of the X-ray low, hard state during the 2008 outburst. During this last outburst, we detected a change of slope in the NIR spectrum between the J and K$_s$ bands, where the JH part is characteristic of an optically thick disk emission, while the HK$_s$ part is typical of optically thin synchrotron emission. Furthermore, the comparison of our IR data with radio and X-ray data shows that $\hdsqt$ exhibits a faint jet both in radio and NIR domains. Finally, we suggest that the companion star is a late-type main sequence star located in the Galactic bulge.}
{These OIR photometric and spectroscopic observations of the microquasar $\hdsqt$, which are the first of this source to be published in a broad multiwavelength context, allow us to unambiguously identify two spectra of different origins in the OIR domain, evolving from optically thick thermal emission to optically thin synchrotron emission toward longer wavelengths. Comparing these OIR observations with other black hole candidates suggests that $\hdsqt$ behaves like a radio-quiet and NIR-dim black hole in the low, hard state. This study will be useful when quantitatively comparing the overall contribution of the compact jet and accretion flow in the energy budget of microquasars.}


\authorrunning{S. Chaty et al.}
\titlerunning{Infrared study of $\hdsqt$ in outburst: a radio-quiet microquasar}

   \maketitle
%

\section{Introduction} \label{section:introduction}

X-ray binary systems are composed of a companion star and a compact object (black hole or neutron star). In low-mass X-ray binaries (LMXBs), the companion star is a late-type star filling its Roche lobe. Matter transiting through the Lagrange point accumulates in an accretion disk around the compact object. LMXBs spend most of their time in a quiescent state with low X-ray luminosity. Outbursts occasionally occur, owing to an instability in the accretion disk,  during which the luminosity increases by several orders of magnitude, especially in the X-rays but also in other wavelengths. Among LMXBs, systems that additionally show nonthermal radio emission, which are sometimes spatially resolved into jets, are called microquasars \cite[see, e.g.,][]{chaty:2007, fender:2006}. During these outbursts, LMXBs mainly exhibit two different stable states, according to their relative X-ray and radio emission and to the presence of QPO (quasi-periodic oscillation): {\it i.} the thermal or high, soft state (HSS) is characterized by a substantial disk fraction ($\geq 75\%$) and the absence of QPO; {\it ii.} the hard or low, hard state (LHS) is recognized by a low disk contribution ($\leq 20\%$) and high fraction of the hard power law component ($\geq 80\%$), as defined in \citet{remillard:2006}.

$\hdsqt$ was discovered during a bright outburst in August 1977 by Ariel V \citep{kaluzienski:1977} and was accurately localized by the High Energy Astronomical Observatory 1 (HEAO1) satellite a few weeks later \citep[1H\,1741-322;][]{doxsey:1977}. This source was probably often active during subsequent years, ever since outbursts were detected in 1984 by EXOSAT \citep{reynolds:1999} and in 1996 by TTM \citep{emelyanov:2000}.
On March 21, 2003, while intensively scanning the Galactic center, {\it INTEGRAL} detected a bright source named IGR\,J17464-3213 \citep[= XTE\,J1746-322;][]{revnivtsev:2003}, which was in fact a rediscovery of $\hdsqt$ \citep{markwardt:2003}.
A bright outburst occurred on April 8, 2003, and a radio counterpart was detected at the Very Large Array \citep[VLA,][]{rupen:2003b}. Relativistic jets from this microquasar were observed with the Australia Telescope Compact Array (ATCA), as interacting with the interstellar medium \citep{corbel:2005}. 

This X-ray binary presents X-ray timing and spectral properties typical of black hole candidates \citep{white:1983}. It exhibited five outbursts between 2003 and 2008, as shown and numbered in Fig.~\ref{figure:XIRlc} (lower panel): 
{\it i.)} an intense outburst in March 2003 (beginning on MJD 52720), followed by four fainter episodes; 
{\it ii.)} in July 2004 (MJD 53200);
{\it iii.)} in August 2005 (MJD 53600); 
{\it iv.)} in December 2007 (MJD 54450); and finally 
{\it v.)} a very faint outburst in September and October 2008 (MJD 54750).
Many X-ray satellites, including {\it INTEGRAL}, {\it Rossi-XTE}, {\it XMM-Newton}, and {\it Swift}, and the radio telescopes VLA and ATCA have followed most of these five outbursts in parallel, allowing a multiwavelength study \cite[see a comprehensive radio/X-ray study of this source from 2003 to 2010 in][]{coriat:2011}. 
Despite more than 35 years of study since the discovery of this microquasar, the nature of both its compact object and the companion star essentially remain unknown.

Here we report on optical and near-infrared (OIR) photometric and spectroscopic observations using European Southern Observatory (ESO) facilities of the counterpart of $\hdsqt$, obtained at the New Technology Telescope (NTT) simultaneously with outbursts {\it i, ii}, and {\it v} (see Fig.~\ref{figure:XIRlc} upper panel). These observations, apart from the R, I, and K$_s$ magnitudes reported in \cite{mcclintock:2009}, are the first OIR observations of this source presented in a broad multiwavelength context. We first describe in Section \ref{section:introduction2} how the three studied outbursts evolved with respect to the various states of the source, in relation to our OIR observations (reported in Table \ref{table:obs}). We then give details on the observations in Section \ref{section:observations}, describe the results, and discuss them in Section \ref{section:results}. We finally conclude in Section \ref{section:conclusion}.

\section{Different outbursts, different states} \label{section:introduction2}

\subsection{The 2003 outburst ({\it i})} 

The {\it INTEGRAL} variable source reported by \citet{revnivtsev:2003} on March 21, 2003, turned out to be an intense outburst from the already known source $\hdsqt$, that lasted several months until November 2003. This outburst, labeled {\it i} in Fig.~\ref{figure:XIRlc}, initially displayed a low, hard X-ray spectrum that gradually softened until a radio flare was reported on April 8, 2003 \citep{rupen:2003c}. The OIR observations \#1 to \#4, reported in Table \ref{table:obs}, were carried out after this radio flare, during a phase that several authors identified as a soft, intermediate state \citep{joinet:2005, homan:2005a, capitanio:2005}.
Nevertheless, a previous work by \citet{parmar:2003} reported three observations with {\it INTEGRAL} on April 6, 14, and 21 and concluded that the source was in the canonical high, soft state at those dates. The spectrum continued evolving according to the typical hardness-intensity diagram for a transient black hole in outburst, beginning with a low, hard state that is followed by hard/intermediate and soft/intermediate states. The system is clearly identified as being in a high, soft state by July and August 2003 \citep{homan:2005a, kretschmar:2003}, after which it made a transition to the hard state around September--October 2003 \citep{grebenev:2003, tomsick:2003b}.

\subsection{The 2004 outburst ({\it ii})}

Another X-ray brightening of $\hdsqt$ was observed with {\it Rossi-XTE} in the beginning of July 2004 \citep{swank:2004}. \citet{rupen:2004a} report VLA radio observations from this system on July 11, 2004, and \citet{rupen:2004b} suggested on August 5, 2004 that the radio rise might be indicative of a transition from high, soft to a harder state. The OIR observations \#5 and \#6 presented in this work (see Table \ref{table:obs}) were taken at the end of July 2004 and suggest that the source was then still in the high, soft state, while the optical observation \#7 was carried out one day after this last radio detection, with the source transiting to a harder state. {\it INTEGRAL} data covered the final part of the outburst and showed that the source, which seemed to harden after the peak of the emission, suddenly fell to a softer state \citep{capitanio:2006}. Our observations \#8 and \#9 correspond to this high, soft state.

\subsection{The 2008 outburst ({\it v})} \label{section:2008}

On September 23, 2008, \citet{kuulkers:2008} reported a new outburst of $\hdsqt$, which displayed a low, hard X-ray spectrum. The system stayed in a low, hard state until October 19, 2008, and began a transition to softer states after that date \citep{prat:2009}. Among the OIR data that we present in this work, the observation \#10 (Table \ref{table:obs}) was performed during the low, hard state of the source. Radio and X-ray observations carried out in October 2008, showed that the source was progressing into a hard/intermediate state \citep{corbel:2008, yamaoka:2008, belloni:2008}, therefore likely to be missing any high, soft state, which might correspond to a ``failed state transition'' \citep{capitanio:2009}.

\begin{figure*}
  \begin{center}
    \includegraphics[angle=90,width=17.cm]{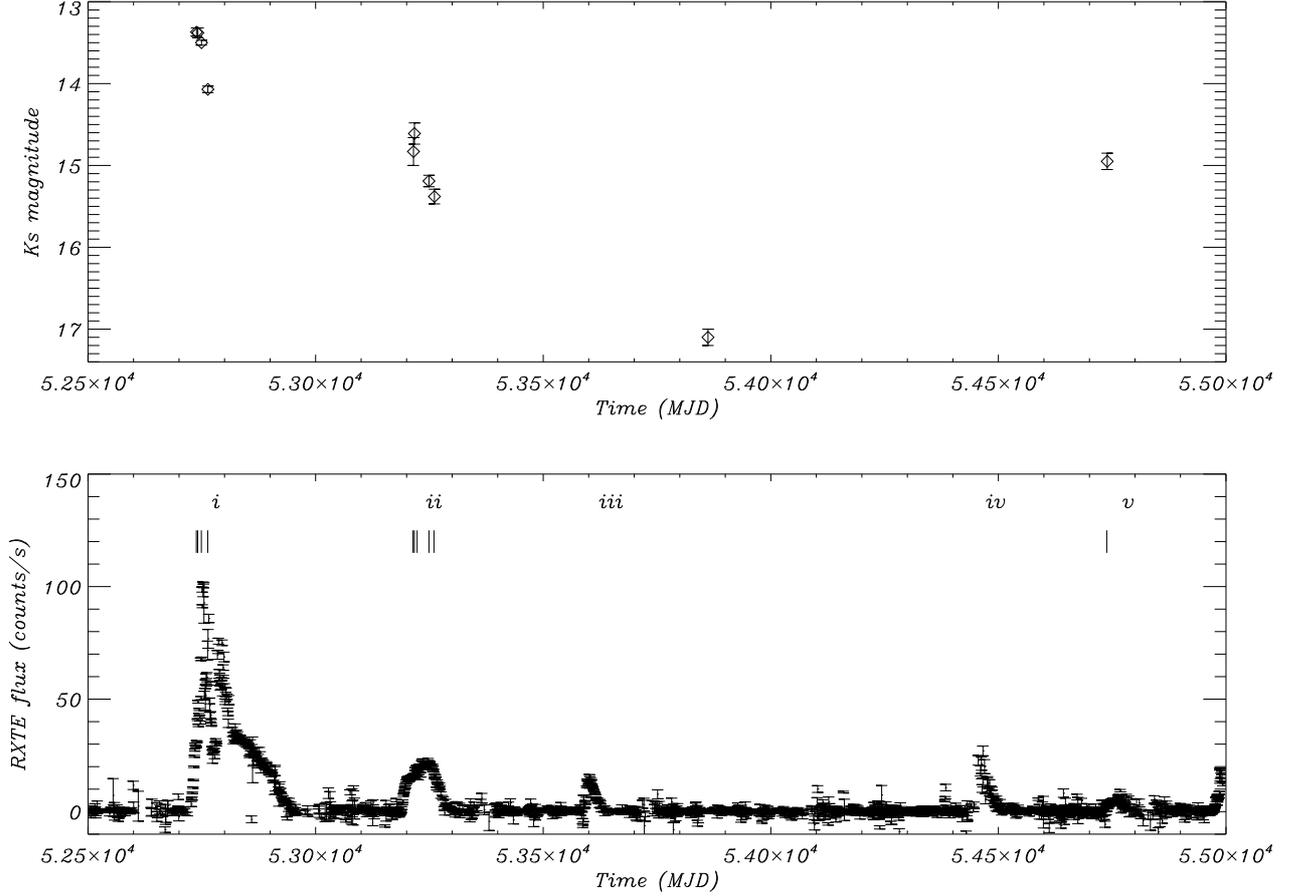}
  \end{center}
\caption{Upper panel: Long-term ESO/NTT NIR K$_s$ lightcurve of $\hdsqt$. We included the quiescent K$_s = 17.1$ magnitude reported by \cite{mcclintock:2009} (MJD 53862). Lower panel: {\it Rossi-XTE}/ASM 2-12 keV X-ray lightcurve. We report in the top of this panel: {\it i.} the five outbursts that occurred from 2003 to 2008 (indicated by numbers, see Section \ref{section:introduction} for details), and {\it ii.} the dates of our OIR observations (indicated by vertical ticks, see Table \ref{table:obs} for details).}
\label{figure:XIRlc} 
\end{figure*}

\section{Observations} \label{section:observations}

\begin{figure}
  \begin{center}
    \includegraphics[angle=0,width=9.cm]{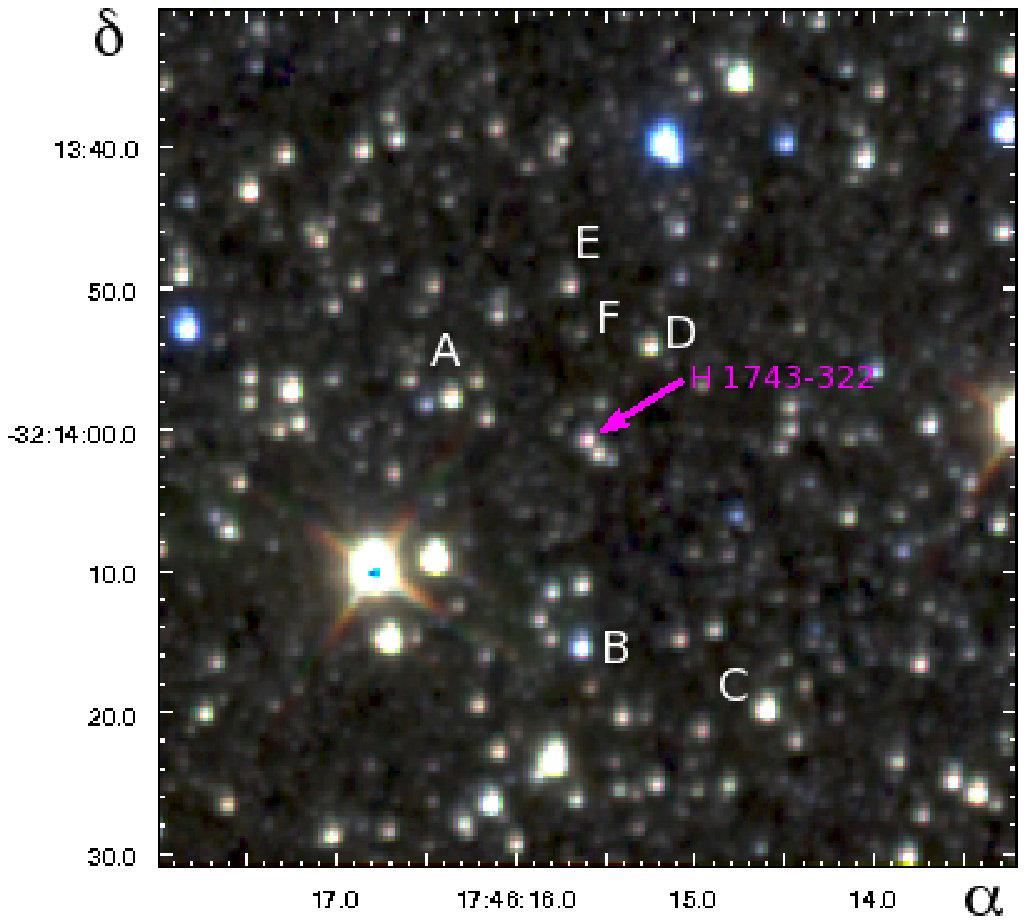}
  \end{center}
\caption{Three-color (J, H, and K$_S$) field of view around the $\hdsqt$ counterpart observed with SOFI on April 9, 2003. The location of the X-ray binary (RA = $17\hour46\min15\secp608$; DEC = $-32\adeg14\amin0\asecp6$ J2000 with $0\asecp5$ uncertainty) is shown by an arrow.} 
\label{figure:field}
\end{figure}

We obtained OIR observations of $\hdsqt$ simultaneously for three outbursts in 2003 (\#{\it i}), 2004 (\#{\it ii}), and 2008 (\#{\it v}), using the {\em European Southern Observatory} (ESO) 3.58m-{\em New Technology Telescope} (NTT) at La Silla, as part of a Target-of-Opportunity (ToO) program (PI S.\,Chaty). Details on the observing dates are given in Table~\ref{table:obs}. Data have been taken with the EMMI, EFOSC2, and SOFI instruments, and encompass OIR photometry, spectroscopy in the $1-2.5 \microns$ range and K$_s$ band polarimetry. Polarimetric data, obtained during observations \#2 and \#9, have already been presented in \cite{dubus:2006b}. A finding chart is shown in Fig.~\ref{figure:field}.
We note the nearby star 1\farcs1 SW of $\hdsqt$ with K$_s=13.65\pm0.03$. 
Stars A-C are used for reference position, and stars C-F for differential photometry: in particular, star C is the flux calibrator (K$_s=11.69\pm 0.03$) for rapid IR photometry, and stars D-F are used as comparisons for $\hdsqt$ (see Section \ref{section:fastIR}).
All data were analyzed with the IRAF (Image Reduction and Analysis Facility) suite \citep{massey:1992}.

\begin{table*}
 \centering
  \caption{Log of OIR observations. JHK$_s$ apparent magnitude photometry of $\hdsqt$ with SOFI. Col.1: spectral state, Col.2: number corresponding to the observation, and the horizontal lines separate the distinct outbursts ({\it i}, {\it ii} and {\it v} respectively). Last Col.: $\Delta$m is the estimated magnitude correction due to blending with the nearby star (see Fig.~\ref{figure:field} and text in Section \ref{section:fastIR}). \label{table:obs}}
  \begin{tabular}{@{}lllllccccccl@{}}
  \hline
State & \# & MJD & Date & Time (UT) & Seeing & R & I & J & H & K$_s$ & $\Delta$m \\
\hline
HSS & 1 & 52738 & 2003-04-09 & 07:15-08:36 & 0\farcs7 & 21.49$\pm$0.05 & 19.48$\pm$0.03 & 15.41$\pm$0.04 & 14.16$\pm$0.07 & 13.37$\pm$0.05 & \\ 
& 2 & 52741 & 2003-04-12 & 07:52-08:06 & 0\farcs6 & & & & & 13.38$\pm0.06$ & 0.05 \\
& 3 & 52749 & 2003-04-20 & 07:54-10:44 & 1\farcs0 & & & & & 13.50$\pm$0.03 & 0.08 \\
& 4 & 52763 & 2003-05-04 & 08:57-10:31 & 0\farcs6 & & & 16.06$\pm$0.06 & 14.83$\pm$0.04 & 14.07$\pm$0.04 & \\
\hline
HSS & 5 & 53215 & 2004-07-29 & 23:21-01:14 & 1\farcs8 & & & & & 14.83$\pm0.17$ & 0.83 \\ 
& 6 & 53217 & 2004-07-31 & 02:07-02:39 & 0\farcs8 & & & 16.62$\pm$0.23 & 15.44$\pm$0.17 & 14.61$\pm$0.13 & \\ 
& 7 & 53223 & 2004-08-06 & 00:36-00:52 & 0\farcs8 & & 20.37$\pm$0.05 & & & & \\
& 8 & 53249 & 2004-09-01 & 00:13-04:05 & 0\farcs8 & & & 17.24$\pm$0.10 & 15.89$\pm$0.05 & 15.19$\pm$0.07 & \\ 
& 9 & 53260 & 2004-09-12 & 23:17-23:31 & 0\farcs6 & & & & & 15.38$\pm0.09$ & 0.18 \\
\hline
LHS & 10 & 54738 & 2008-09-30 & 00:15-01:20 & 0\farcs8 & & & 17.95$\pm$0.10 & 16.45$\pm$0.08 & 14.95$\pm$0.06 & \\ 
\hline
\end{tabular}
\end{table*}

\subsection{Optical BVRIZ photometry}

We observed the field in BVRIZ filters with the EMMI instrument (ESO Multi-Mode Instrument\footnote{http://www.ls.eso.org/sci/facilities/lasilla/instruments/emmi/current Manual/EMMI\_manual\_current.pdf}) on the NTT in two epochs (\#1 and \#7). Data reduction was standard, involving dark correction, flat fielding, and elimination of bad pixels. We obtained a set of photometric standard stars --Mark\,A3, SA\,110\,362, SA\,110\,364, and SA\,113\,177-- \citep{landolt:1992} to flux-calibrate the images. The counterpart of $\hdsqt$ was detected with the magnitudes reported in Table \ref{table:obs}. We also took rapid photometry in the V filter, however the signal-to-noise ratio (S/N) is too low to detect any significant variation.

\subsection{Infrared JHK$_s$ photometry}

IR photometric observations were taken during the three outbursts covered with the IR spectroimager SOFI \citep[Son Of ISAAC\footnote{http://www.eso.org/sci/facilities/lasilla/instruments/sofi/doc/ manual/sofiman\_2p20.pdf};][]{lidman:1999}, using the large field of view $4\aminp92 \times 4\aminp92$, $0\asecp288$ per pixel in J ($1.247 \pm 0.290 \microns$), H ($1.653 \pm 0.297 \microns$), and K$_s$ ($2.162 \pm 0.275 \microns$) filters. We used various dithering offsets to allow a good determination of the thermal background. 
We also observed the photometric standard stars sj9155, sj9172, and sj9181 in 2003, sj9175 on July 31, 2004, and sj9183 on September 1, 2004 \citep{persson:1998}.
The mean zero points we derived were 1.915 in J, 2.112 in H and 2.699 in K$_s$, with an airmass between 1.085 and 1.181.

Data reduction was standard: after flat fielding, bad pixels were flagged and the background subtracted using a sky map created from the median of the ten dithered frames offset from the target position and bracketing each exposure. Frames were then cross-correlated with a reference exposure to obtain the shift to subpixel accuracy. Positions for $\hdsqt$ and several other stars in the field were determined from the reference frame and taken as input for the aperture photometry. We used the IDL astronomy user's library\footnote{http://idlastro.gsfc.nasa.gov/homepage.html} photometry routines based on the {\it daophot} package \citep{landsman:1993} to derive apparent magnitudes, especially the crowded field photometry tasks, due to the high level of blending of $\hdsqt$, taking the aperture correction into account.
We computed the astrometry of the fields in the three filters using the {\it gaia} software, taking the position of more than 1000 objects from the 2MASS catalog into account. We reached an accuracy of $0\asecp209$ per pixel.

We report the apparent magnitudes in Table \ref{table:obs}, where the errors are dominated by the final aperture correction, in most of the cases less than $0.1 \mags$. We also checked our final photometry by verifying that the star close to $\hdsqt$ (see Fig.~\ref{figure:field}) was not variable in our error range of $0.1 \mags$, and additionnally more than 50\% of the stars in the field of view do not vary within this range.

\begin{figure*}
	\begin{center}
	\includegraphics[angle=90, width=8.9cm]{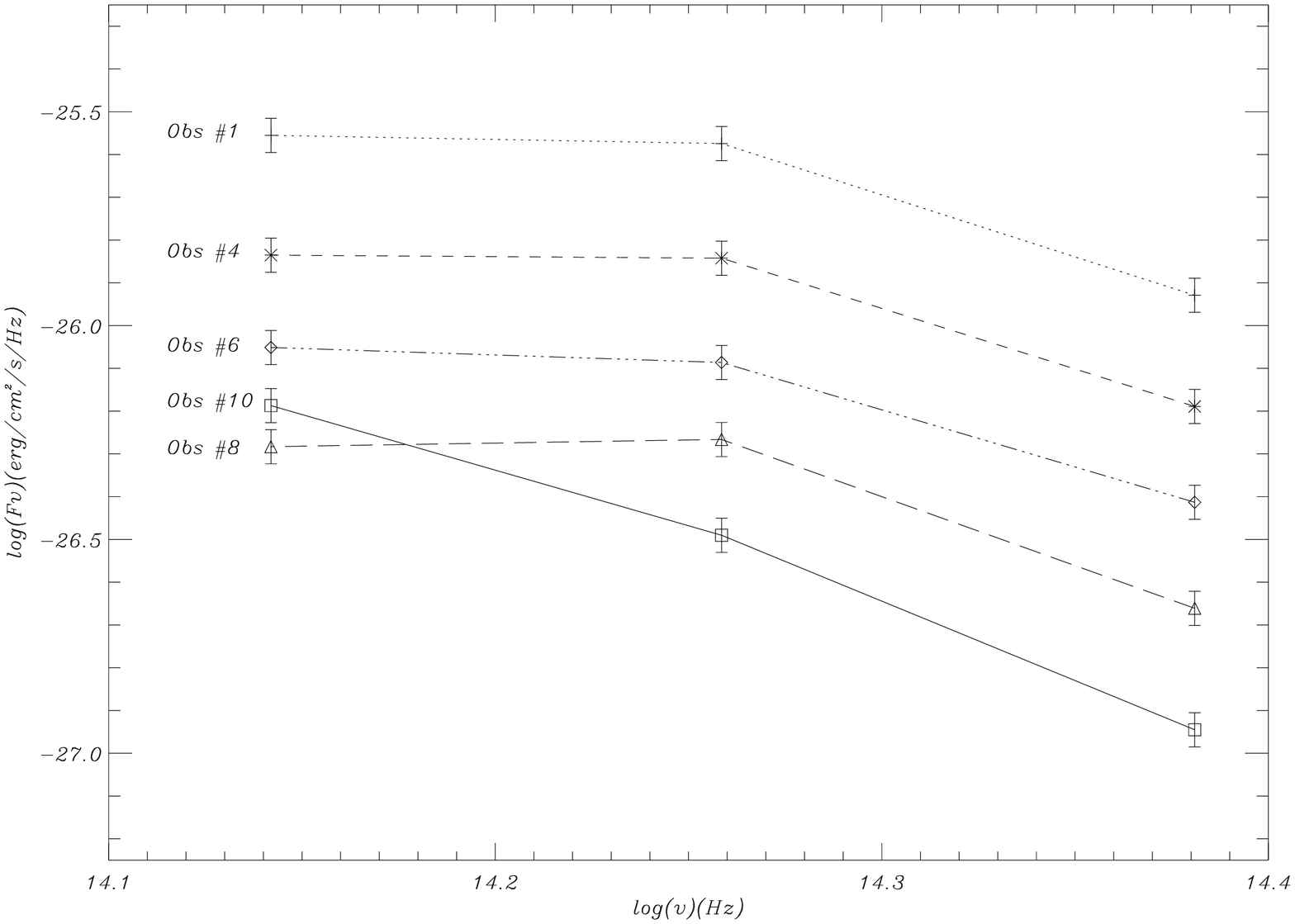}
	\includegraphics[angle=90, width=8.9cm]{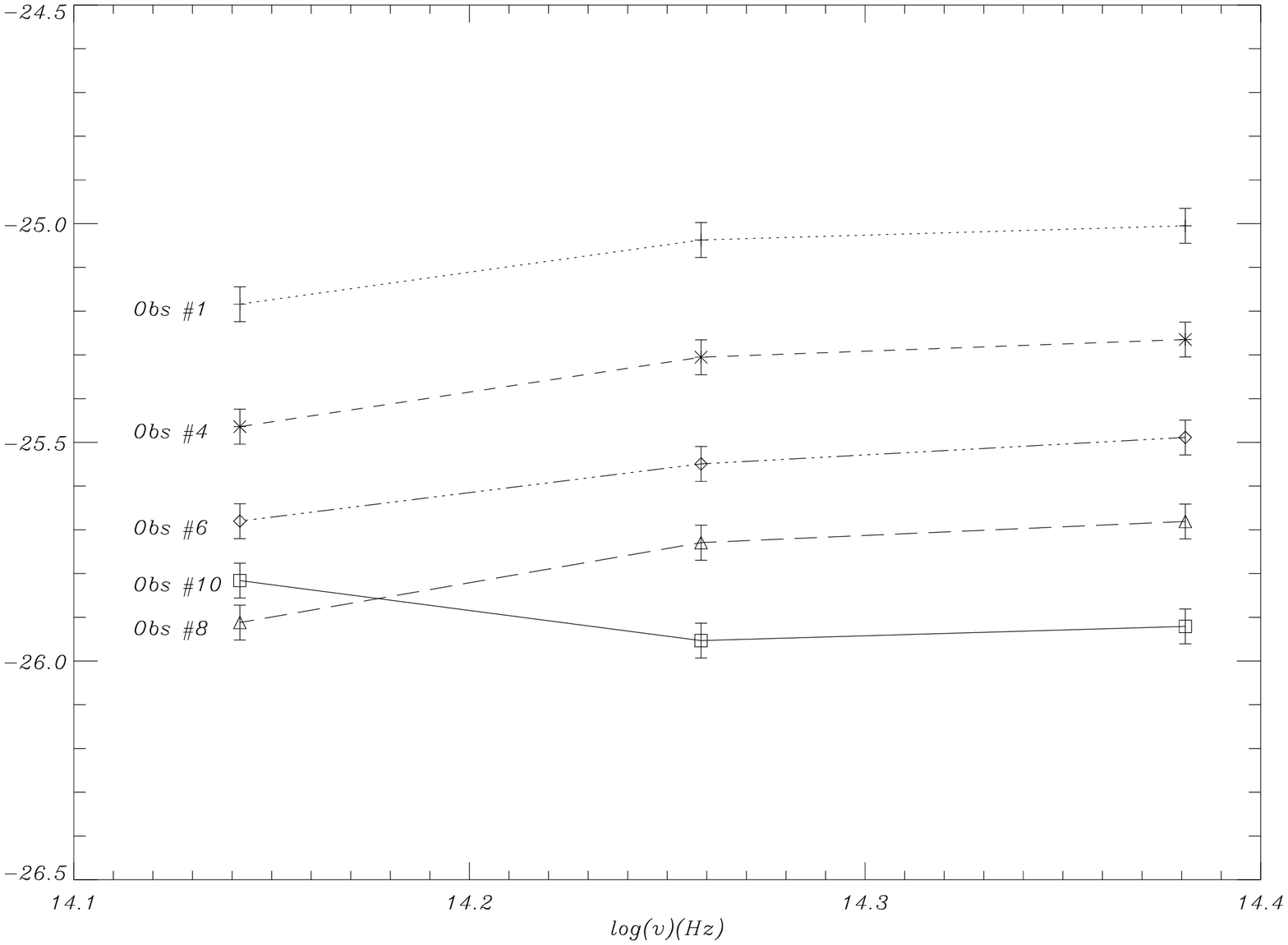}
	\end{center}
\caption{JHK$_s$ photometry of $\hdsqt$. We show the spectral energy distribution, with reddened magnitudes on the left, and dereddened on the right (corrected from interstellar absorption, see text for details). While the y-axis is offset between the two panels, it has the same scale to better show the effect of the dereddening.
}
\label{figure:SED-IR} 
\end{figure*}

The results of JHK$_s$ photometry are shown in Figures\,\ref{figure:XIRlc} (upper panel) and \ref{figure:SED-IR}, where we built the spectral energy distribution (SED) on different epochs, with reddened magnitudes in the left hand column and dereddened ones in the right hand one. Apparent magnitudes have been dereddened considering the absorption on the line of sight, derived from: the hydrogen column density coming from {\it Swift} and {\it XMM-Newton} observations $\nh = 1.8 \pm 0.2 \times 10^{22} \cmmoinsdeux$ \citep{prat:2009}; the relationships $\Av = 4.52 \pm 0.18 \times 10^{-22} \nh$ \citep{guver:2009} and $R_v = \Av / E(B-V) = 3.1$ \citep{cardelli:1989}; and the conversion $\Av$ -- $A_\lambda$ for various filters \citep{indebetouw:2005}. This leads to $\Av = 8.14 \pm 0.15 $, $\Aj = 2.31 \pm 0.15$, $\Ah = 1.43 \pm 0.10$, and $\Ak = 0.93 \pm 0.05 \mags$.


We note that dereddening the magnitudes with higher absorption leads to spectral indices that are not compatible with thermal emission in the soft state. This suggests that the column density we use here \citep[$\nh = 1.8 \times 10^{22} \cmmoinsdeux$,][]{prat:2009} seems to be more reliable than higher ones, such as $\nh = 2.3 \times 10^{22} \cmmoinsdeux$ \citep{miller:2006} or $\nh = 2.5 \times 10^{22} \cmmoinsdeux$ \citep{parmar:2003}.

\subsection{Infrared K$_s$ rapid photometry} \label{section:fastIR}

We also obtained rapid K$_s$ photometry (2s time resolution) of $\hdsqt$ during observations \#3 to \#5, and \#8. Since the target is only 1\farcs1 away to the NE of the nearest star (see Fig.~\ref{figure:field}), that causes centroiding errors when the source is faint. To remedy this problem, we measured the offset of $\hdsqt$ from three nearby bright stars (stars A-C in Fig.~\ref{figure:field}) using the summed image of May 4, 2003, which has the best seeing. The centroid of each reference star was then found in each exposure and the $\hdsqt$ aperture placed at the predetermined offset. We chose an aperture radius of two pixels ($\approx 0\farcs6$) as a compromise between maximizing the S/N and minimizing the contamination from the nearby star to $\hdsqt$. The sky scatter was estimated from an annulus 10-30 pixels away from the centroid and added in quadrature to the Poisson standard deviation. We averaged the three renormalized lightcurves to get the final differential photometry. Lightcurves were also constructed for stars D, E, and F using this procedure.

\begin{figure*}
	\begin{center}
	\includegraphics[angle=0, width=18cm]{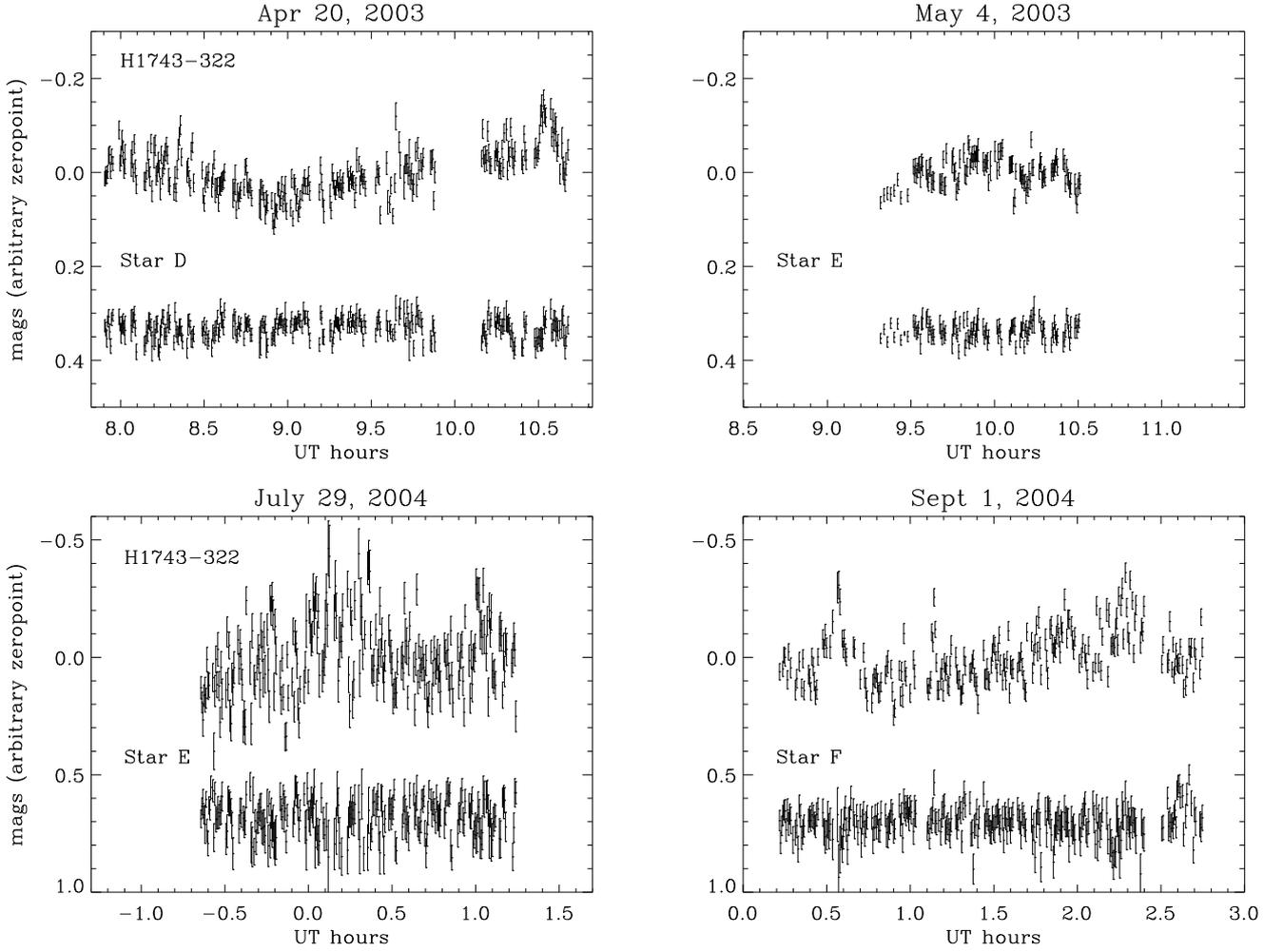}
	\end{center}
\caption{Rapid K$_s$ photometry of $\hdsqt$ for observations \#3, 4, 5, and 8. Each point is a 2\,s integration. We also show a comparison star of similar brightness in each panel. The y-axis is in magnitudes with arbitrary offsets (note the larger y scale in both lower panels).}
\label{figure:lightc}
\end{figure*}

The count rates were converted to magnitudes using star C as a reference. This is the brightest, most isolated, and unsaturated star in Fig.~\ref{figure:field}. Its K$_s$ band magnitude was found to be $11.69 \pm 0.03$ by comparison with photometric standards (sj9155 on April 9, 2003 and sj9172 on May 4, 2003) from \citet{persson:1998}. The growth curve (i.e. how the flux increases with aperture size) for each frame of stars A-F matches that of star C very well, so our estimate of the magnitudes should be reliable even when using a small aperture. Indeed, the lightcurves for our check stars, D, E, and F, were constant (Fig.~\ref{figure:lightc}) and their estimated magnitude (respectively K$_s=13.14$, $13.79$ and $15.13$) consistent from one set of observations to the other within $\pm 0.03 \mags$.

The nearby star can contribute a significant amount of flux in the $\hdsqt$ aperture (and vice-versa) when the seeing is poor. To evaluate the amount of blending, we placed two regions around star D at offsets corresponding to the $\hdsqt$ - nearby star vector. These regions were used to obtain an image by image estimation of the contamination in each aperture. The magnitude correction $\Delta$m applied to the raw $\hdsqt$ flux is given in Table~\ref{table:obs}. The final magnitude given is the magnitude after correction. The error does not take uncertainties in the deblending into account. An analogous correction was applied to the nearby star, and we recovered the average magnitude of the nearby star (K$_s=13.65\pm0.03$) to within $1\sigma$ errors in all the sets.

\subsection{Infrared blue and red grisms spectroscopy}

We obtained IR blue ($0.95-1.64 \microns$) and red ($1.53-2.52 \microns$) grisms, $R=1000$, spectroscopy with the SOFI instrument, using a $1 \times 290 \asec$ slit, inclined at $45\adeg$, on four different dates:
April 19, 2003 06:55-07:50; April 21, 2003 07:06-08:20; July 31, 2004 03:09-04:24; and September 1, 2004 04:16-05:31. We also obtained spectra from the spectro-photometric standard stars Hip\,89321 and Hip\,87810.

\begin{figure*}
  \begin{center}
    \includegraphics[angle=0,width=9cm]{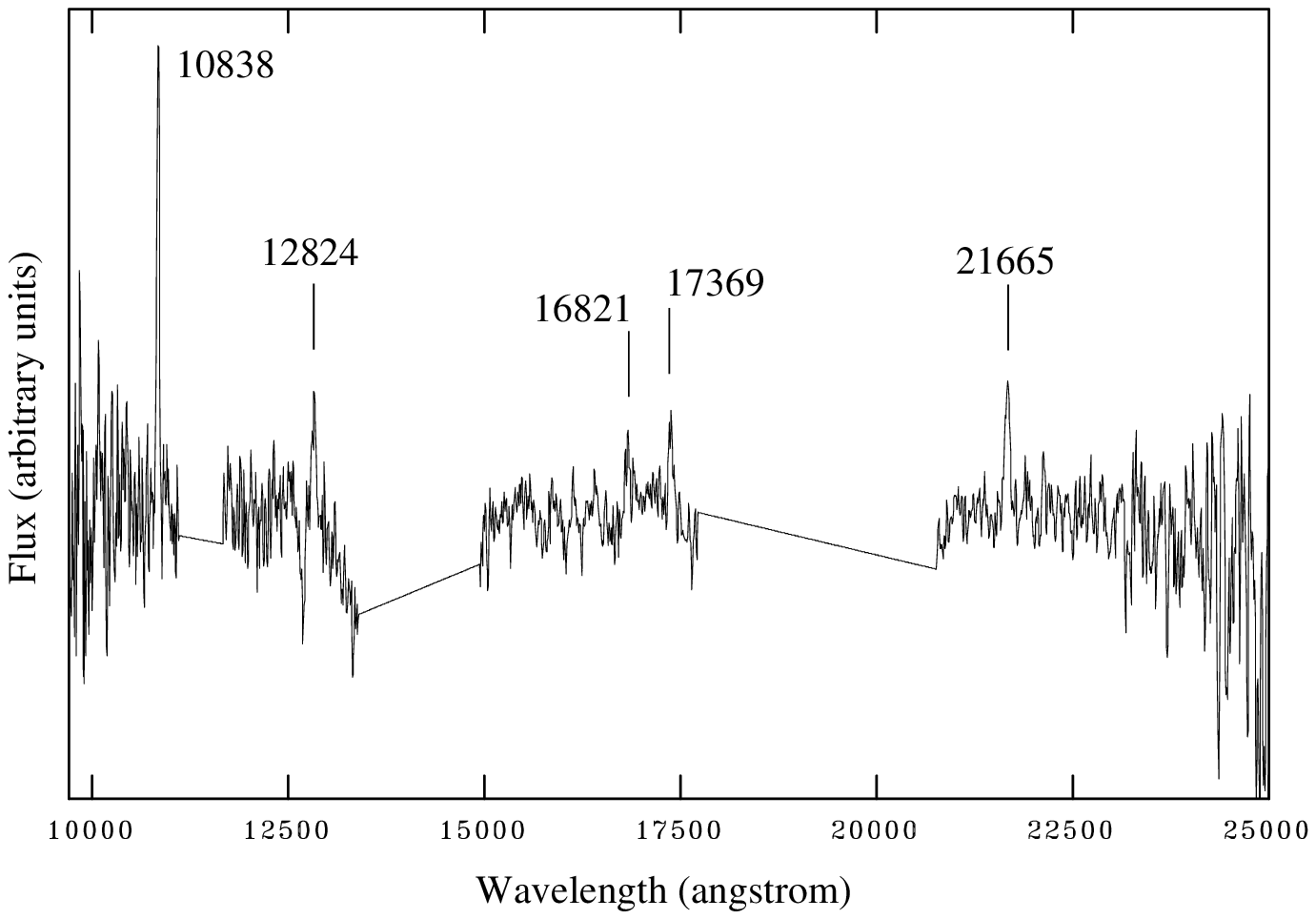}
    \includegraphics[angle=0,width=8.9cm]{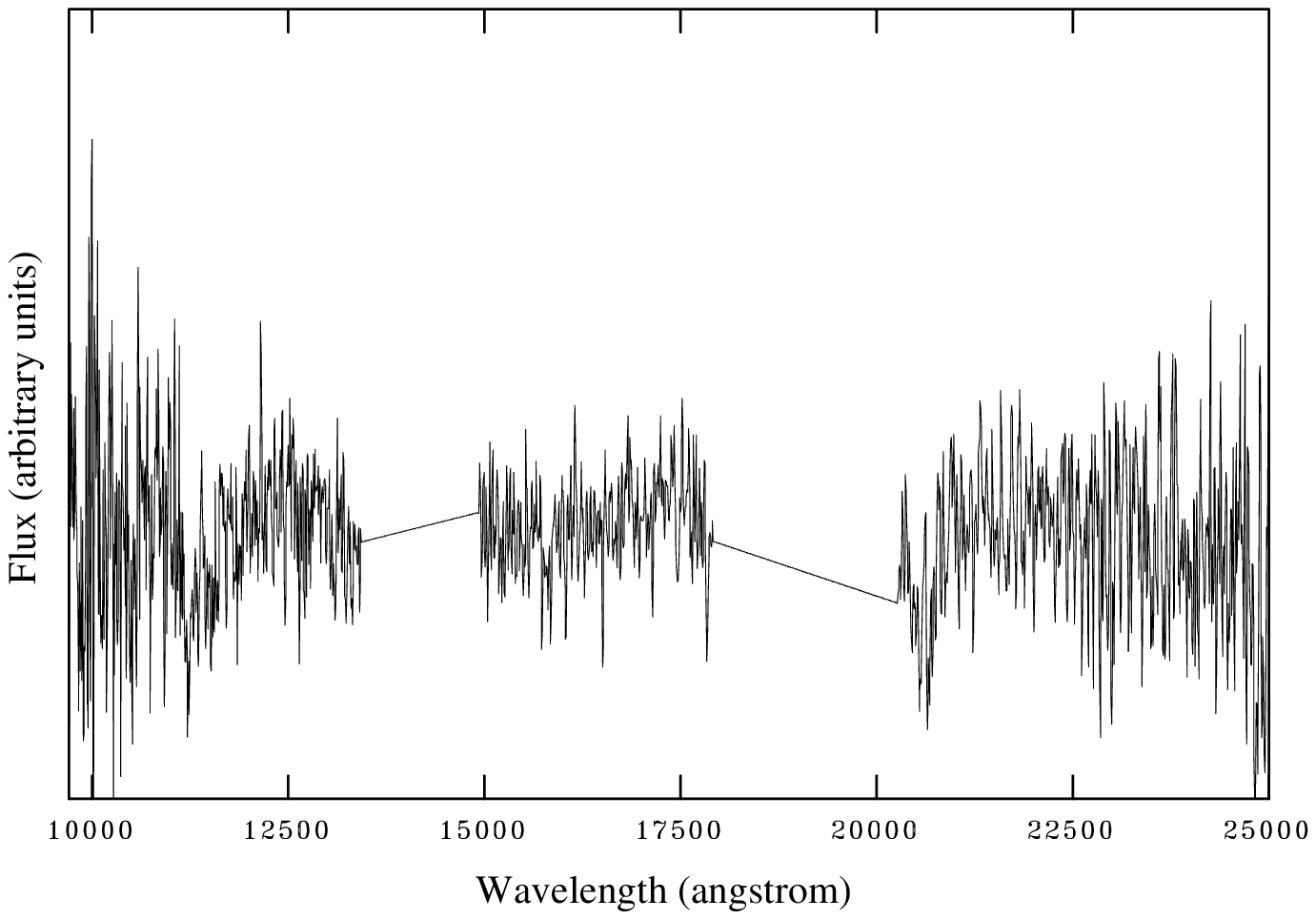}
  \end{center}
\caption{IR SOFI blue and red grism spectroscopy of $\hdsqt$ on April 19, 2003  (left) and July 31, 2004 (right). Both parts around 1.3 and 1.8 $\mu$m are not shown due to high level of atmospheric absorption. We indicate the position of the lines that we could detect and identify, reported in Table \ref{table:lines}.}
\label{figure:spectra}
\end{figure*}

We extracted, wavelength-calibrated with Xenon lamp spectra, median-combined, corrected for telluric absorption, normalized ({\it continuum} task), and finally combined both blue and red grism spectra ({\it scombine} task), reported in Fig.~\ref{figure:spectra} (spectrum taken on April 19, 2003 on the left, and on July 31, 2004 on the right).
In the spectra taken in 2004, the continuum was detected without any clear emission line (the source being faint, the S/N is quite low). On the other hand, in the spectra taken in 2003 (when the source was brighter), a few emission lines are clearly detected: Bracket-$\gamma$, Paschen-$\beta$, and ionized helium lines, reported in Table \ref{table:lines}.

\begin{table*}
 \centering
  \caption{Lines detected in $\hdsqt$ IR spectra with SOFI (wavelength taken from UKIRT database). \label{table:lines}}
  \begin{tabular}{ccccc}
  \hline
Wavelength fit (nm) & Wavelength lab (nm) & FWHM (nm) & Identification & Date \\ 
\hline
1083.8 & 1083.3 & 4.2 $\pm 0.1$ & HeI 2P2S, 3Po3S & 2003-04-19 \\
1084.0 & ''     & 2.9 $\pm 0.1$ &''              & 2003-04-21 \\
1282.4 & 1282.2 & 5.8 $\pm 0.1$ &H (Paschen) & 2003-04-19 \\
1282.7 & ''     & 4.1 $\pm 0.1$ &''          & 2003-04-21 \\
1682.1 & 1681.1 & 6.9 $\pm 0.1$ &H (Brackett) & 2003-04-19 \\
1682.8 & ''     & 6.8 $\pm 0.1$ &''           & 2003-04-21 \\
1736.9 & 1736.7 & 5.9 $\pm 0.1$ &H (Brackett) & 2003-04-19 \\
1737.1 & ''     & 3.2 $\pm 0.1$ &''           & 2003-04-21 \\
2166.5 & 2166.1 & 7.1 $\pm 0.1$ &H (Brackett) (or HeI-2166.0) & 2003-04-19 \\
2166.6 & ''     & 5.0 $\pm 0.1$ &''                           & 2003-04-21 \\
\hline
\end{tabular}
\end{table*}

\section{Results and discussion}  \label{section:results}

\subsection{Variability in IR and X-ray lightcurve}

With the detection of a high level of variation --\, on both long and short timescales, see Figures\,\ref{figure:XIRlc} (upper panel) and \ref{figure:lightc}, respectively\,-- we confirm that the varying IR source is the counterpart of $\hdsqt$. The long-term K$_s$ band IR lightcurve is well correlated with the X-ray lightcurve, with IR emission increasing and decreasing at each X-ray outburst in 2003 and 2004 (observations taken during HSS, see Section \ref{section:introduction2}). On the other hand, the IR observation obtained in 2008, while the source was in LHS is decorrelated from the overall X-ray lightcurve. This is consistent with the IR vs X-ray correlation described in \citet{russell:2006a}.

The short-term IR lightcurve (Fig.~\ref{figure:lightc}) shows a flaring modulation on sub-hour timescale, possibly due to orbital period or even superhump period. The coverage is, however, not long enough to tell, and Lomb-normalized periodogram (LNP) gives nothing. The level of short-term variability of $\hdsqt$ relative to comparison stars is higher in 2004 than 2003 by a factor 2, which is consistent with the fact that the source is transiting to the LHS during observation \# 7 (2004 outburst {\it ii}), while it is in the canonical HSS in observations \#3 and \#4 (2003 outburst {\it i}, see Section \ref{section:introduction2}). 

This lightcurve presents similarities with the one of $\gx$ obtained by \cite{casella:2010} during a low, hard state. In both cases there are flares, however they seem to typically last longer for $\hdsqt$ ($\sim 10$\,minutes for 2004 Sept. 1 observation \#8) than for $\gx$ ($\sim 5$\,minutes, see their Figure 1). We note that $\gx$ also exhibits faster variability \citep{casella:2010}, which is outside of our detection possibility for $\hdsqt$.
This indicates that the short-term variations in 2004 (outburst {\it ii}) are likely due to nonthermal processes at the base of the compact jet near the black hole \cite[see also][]{chaty:2011a}, while the emission is purely thermal, coming from the accretion disk, in 2003 (outburst {\it i}).

We point out that the spectra taken in 2003 and reported in Fig.~\ref{figure:spectra} (left panel) exhibit H and He emission lines most likely emanating from the accretion disk, which is consistent with the picture above, of OIR observations of 2003 corresponding to the HSS. Alternatively, those taken in 2004 (Fig.~\ref{figure:spectra} right panel) are featureless, corresponding to the source transiting to the LHS.

Finally, during the short outburst ({\it v}) between October 3 and November 16, 2008, the source made a transition from the LHS to the hard, intermediate state, and then decreased in luminosity while its spectrum hardened \citep{capitanio:2009}. Since $\hdsqt$ did not follow the canonical pattern through all the spectral states (see Section \ref{section:2008}), \cite{capitanio:2009} propose that the source exhibited a kind of ``failed'' outburst, rarely observed before in a transient black hole candidate. This might explain both the low-level and distinct spectral index in IR of the source, as seen in our OIR observations (see Fig.\,\ref{figure:SED-IR}). Compact jets have been detected in radio during this type of outburst, for instance in the microquasar $\xtejqcq$ \citep[see][and discussion thereafter]{chaty:2011a}, with a contribution of their emission in the IR domain.

\subsection{Multiwavelength SED} \label{section:SED}

\subsubsection{IR SED}

In the IR SED reported in Figure \ref{figure:SED-IR}, we clearly see that the maximum source flux decreases along the years from 2003 to 2008 outbursts, as shown in the X-ray lightcurve (Fig.~\ref{figure:XIRlc}), similarly to an emptying reservoir which would be constituted by the accretion disk. During the 2003 and 2004 outbursts (observations \#1, 4, 6, and 8), the SED is nearly flat, with dereddened IR spectral indices\footnote{The spectral index $\alpha$ is defined as $F_{\nu} \propto \nu^{\alpha}$.}, reported in Table \ref{table:IS}, characteristic of optically thick thermal emission coming from the accretion disk ($\alpha_{JH} \sim 0$). However in 2008 (observation \#10), the SED clearly shows that the spectral slope is inverted in the NIR domain. While the spectral index between J and H ($\alpha = 0.0$) is typical of an optically thick regime, the one between H and K$_s$ ($\alpha = -0.9$) is inverted, which is characteristic of synchrotron emission in an optically thin regime. Even if very steep, this HK$_s$ spectral index is similar to the ones of black-hole X-ray binaries in LHS, such as $\alpha = -0.7$ for $\gx$ \cite[see e.g. ][]{gandhi:2011,rahoui:2012,russell:2013}.

We also note from Figure \ref{figure:SED-IR} that the source contributes more to K$_s$ flux compared to J flux in 2008 than during other epochs, which is consistent with LHS emission. This points strongly towards a link between long-wavelength IR emission and a jet component \citep{russell:2006a}.

\subsubsection{Radio-IR-X-ray SED}

\begin{figure}
	\begin{center}
	\includegraphics[angle=90, width=8.9cm]{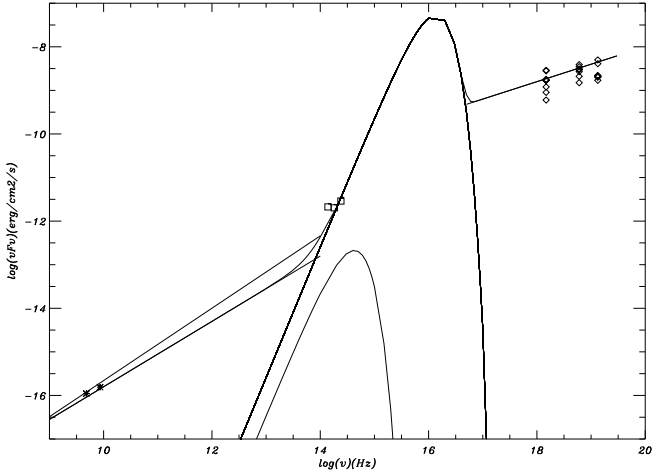}
	\end{center}
\caption{Broad-band multiwavelength SED of $\hdsqt$ for observation \#10, while the source is in the LHS, including dereddened IR magnitudes, radio and X-ray fluxes, a stellar emission, and an accretion-disk model. We also plot the minimum and maximum radio slopes based on flux error bars, in order to extrapolate the radio flux towards the IR (see details in Section \ref{section:SED}).}
\label{figure:SED-broad} 
\end{figure}

To disentangle the various contributions emitted by this microquasar, we show in Fig.~\ref{figure:SED-broad} the broad-band multiwavelength SED of $\hdsqt$ for the observation \#10 (2008 outburst {\it v} in LHS), including dereddened IR magnitudes, radio and X-ray fluxes, a model of stellar emission corresponding to a G0\,V star of temperature 5200\,K and of solar radius, and an accretion disk model in a typical low, hard state, which was adjusted to fit the radio and X-ray data contemporaneous\footnote{While data are not strictly simultaneous, we took the most close-in-time existing data.} to our OIR observations. We included {\it i.} ATCA radio flux densities of $2.30 \pm 0.10$\,mJy at 4.8\,GHz and $1.80 \pm 0.15$\,mJy at 8.6\,GHz, corresponding to a spectral index of $-0.42 \pm 0.16$ \citep[data taken on October 5, 2008,][]{corbel:2008}, and {\it ii. Rossi-XTE}--ASM (2-12 keV), PCA (4-45 keV), and HEXTE (20-100 keV)-- X-ray data from the 2008 October outburst \citep{capitanio:2009}.

We immediately see that, while the JH part of IR emission is consistent with (optically thick) thermal emission from the accretion disk, the HK$_s$ part is clearly inverted, therefore not thermal, that it has a (optically thin) synchrotron nature, most likely due to a contribution from the compact jet. It is interesting to notice that the V-shape break of the SED during observation \#10 (Figures \ref{figure:SED-IR} and \ref{figure:SED-broad}) is reminiscent of the SED of $\gx$ in the low, hard state \cite[see for instance Figure 4 in][]{homan:2005b}. In both cases, it is likely that the optically thick spectrum (JH, $\alpha \sim 0$) mainly emanates from the accretion disk and that the optically thin synchrotron spectrum (HK$_s$, $\alpha = -0.9$) is produced by the jet. 

The spectrum then evolves from a very steep HK$_s$ slope towards an inverted radio spectrum, with the spectral break from optically thick to optically thin emission occurring at lower frequencies than from the K$_s$-band. We clearly see in Fig.\,\ref{figure:SED-broad} that the radio spectrum, extrapolated towards the OIR domain, underpredicts the IR K$_s$-band flux (as shown by the minimum and maximum radio slopes based on flux error bars). Of course, we have to be cautious, first because of the error on the radio spectral index derived from only two data points \citep[see e.g. Fig.~1 in][]{brocksopp:2004} and second because the radio data were not taken simultaneously with the IR, but five days later. However, a complex radio-to-IR spectrum, differing from a simple broken power law, has already been reported, for instance, in $\gx$ \citep{corbel:2013}, where mid-IR data are not located on a straight line between radio and IR data, rather suggesting a complex radio-to-IR evolution. Such a complex radio-to-IR SED may be explained by the fact that IR photons emanate from the base of the jet, while radio photons come from further away inside the jets \cite[see also ][about $\hdsqt$]{miller-jones:2012}. 

\subsubsection{OIR vs X-ray correlation}

\begin{figure}
	\begin{center}
	\includegraphics[angle=0, width=10.2cm]{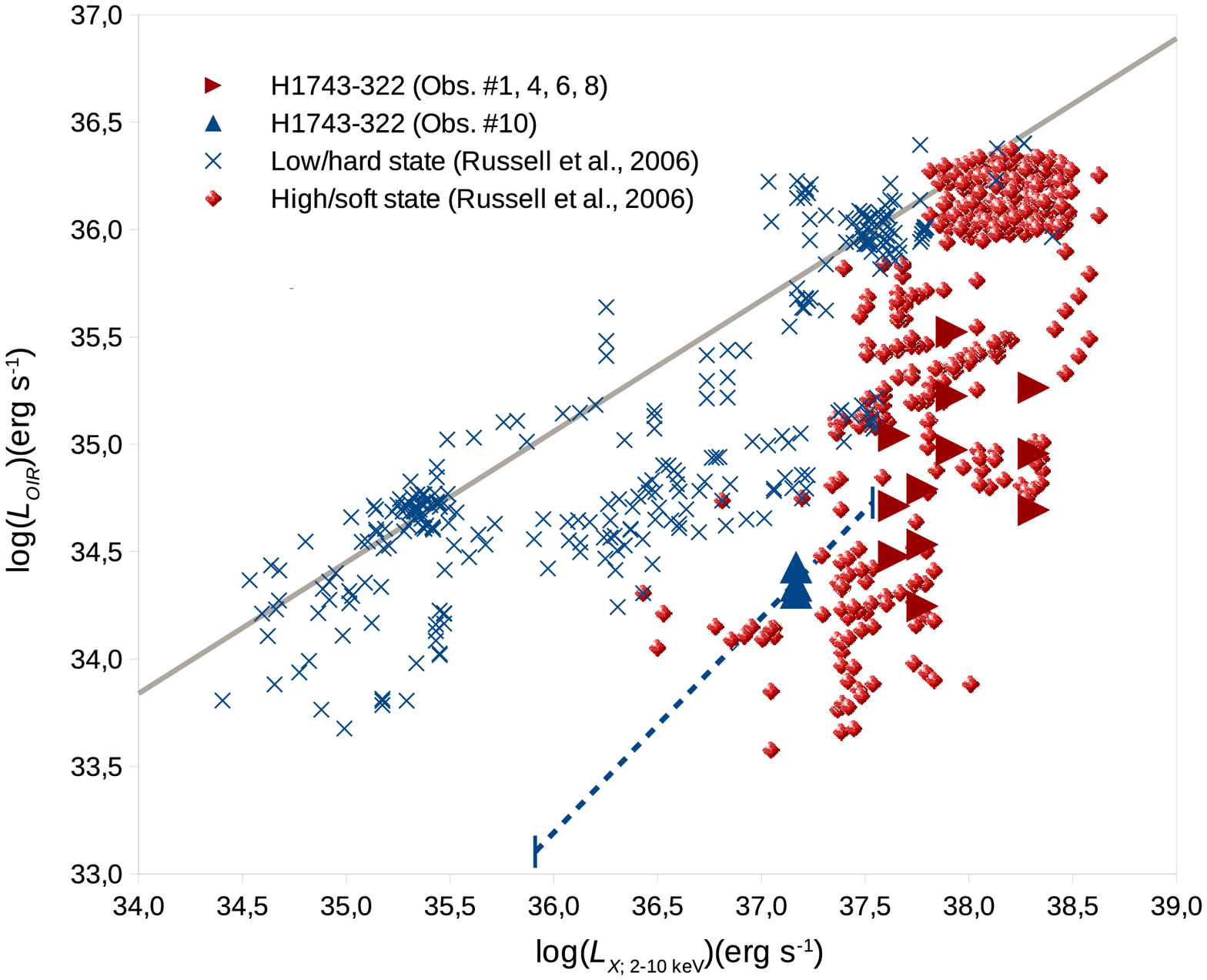}
	\end{center}
\caption{OIR vs X-ray correlation for low-mass black-hole X-ray binaries in low, hard and high, soft states, with all data points taken from \cite{russell:2006a}. The central line represents the best power law fit for low, hard data points: $L_{OIR} = 10^{13.1\pm0.6} L_X^{0.61\pm0.02}$. We added the J, H, and K$_s$ data points of $\hdsqt$ reported in this paper (observations \# 1, 4, 6, and 8 corresponding to HSS and observation \# 10 corresponding to LHS, respectively), computed at the distance of 8.5 kpc. The dashed line corresponds to the different positions of Obs. \#10 (triangle) in LHS, computed for distances between 2 and 13 kpc. We see that while HSS data points of $\hdsqt$ are consistent with HSS data points of other sources, LHS data points of $\hdsqt$ are clearly located below the LHS data points of other sources, precisely between LHS and HSS data points of other sources.}
\label{figure:correlation}

\end{figure}

When adding our OIR data points of $\hdsqt$\footnote{We assume here a distance of 8.5\,kpc for $\hdsqt$.} to the IR vs X-ray correlation given in \citet{russell:2006a}, for which there was no data of $\hdsqt$, we find that $\hdsqt$ HSS data points are consistent with other sources, while LHS data points are clearly located below the LHS correlation for other sources, in fact precisely in between LHS and HSS data points of other sources (see Fig.\,\ref{figure:correlation}). This is consistent with its ``outlier'' position in the radio vs X-ray correlation, which \cite{coriat:2011} suggest is due either to radiatively efficient accretion flow that produces the X-ray emission in the LHS or to radiatively inefficient accretion coupled with unusual outflow properties of the source. We thus notice that the ``outlier'' position has more to do with a faint radio flux than with a high X-ray flux, which is quite typical of black hole candidates in the LHS.

Our observation of the break in the IR domain is interesting in this ``outlier'' context. Indeed, {\it i.)} the JH part fits well with thermal emission from the accretion disk, {\it ii.)} the X-ray properties of $\hdsqt$ are typical of black hole X-ray binaries in LHS, and {\it iii.)} the HK$_s$ part does not extrapolate to the observed radio fluxes. All these facts suggest that the outlier position of $\hdsqt$ is not due to a radiatively efficient accretion flow, but instead to unusual outflow properties, which \cite{coriat:2011} suggest could be a linear dependence between the jet power and the accretion rate. This is also consistent with the fact that, as reported by \citet[][Fig.\,4 lower panel]{russell:2013}, the radio-faint black hole X-ray binaries exhibit lower jet luminosity at the break frequency, compared to radio-loud ones. In other terms, our IR observations show that $\hdsqt$, instead of being X-ray loud, would rather be a radio-quiet and NIR-dim black hole in the LHS, and this would constitute, after the microquasar $\xtejscq$ \citep{curran:2012}, the second clear example of an outlier in the NIR vs X-ray correlation.

\begin{table}
 \centering
  \caption{Spectral indices of JH, HK$_s$, and JK$_s$ power laws (data taken from Figure \ref{figure:SED-IR} right panel). The typical error on the spectral indices is $\sim 0.10$. \label{table:IS}}
  \begin{tabular}{crrr}
  \hline
Observation & JH & HK$_s$ & JK$_s$ \\ 
\hline
\#1  &  0.0 &  1.6 &  0.7 \\
\#4  &  0.0 &  1.7 &  0.8 \\
\#6  &  0.2 &  1.4 &  0.8 \\
\#8  & -0.3 &  1.9 &  0.7 \\
\#10 &  0.0 & -0.9 & -0.4 \\
\hline
\end{tabular}
\end{table}

\subsection{The nature of the companion star}

While the nature of the companion star of $\hdsqt$ is not known yet, we can use its quiescent K$_s$ magnitude of 17.1 \cite{mcclintock:2009} to compute an absolute magnitude of K$ \sim 2$ (using mean column densities derived for the X-ray source). Then, the comparison with catalogued spectral types of stars in our Galaxy \citep[see e.g.][]{ruelas-mayorga:1991} leads to the suggestion that the companion star is a late-type main sequence star located in the Galactic bulge.

\subsection{Comparison with $\xtejqcq$}

The results of this OIR study are somehow reminiscent of the microquasar $\xtejqcq$. The long-term X-ray behavior of both microquasars, which displays several X-ray outbursts, presents many similarities. The X-ray lightcurve of $\xtejqcq$ first shows an intense outburst of the order of 500\,{\it Rossi-XTE}/ASM\,counts\,s$^{-1}$, which took place in 1998 and lasted for $\sim 200$\,days, comprising all the canonical X-ray spectral states. It was followed by a less powerful outburst in 2000, which also exhibited several spectral states, and three much weaker outbursts in the period 2001--2003. 

The OIR observations of $\xtejqcq$ described in \citet{chaty:2011a} were taken simultaneously to its 2003 faint outburst, with the source in its low, hard state. This can be compared to the faint X-ray outburst of $\hdsqt$ observed in 2008 ({\it v}) during its low, hard state. The most evident similarity is related to the SED of both microquasars, the IR photometric results pointing to a nonthermal inverted spectra in both cases. (Compare observation \#10 of $\hdsqt$ in Figs.~\ref{figure:SED-IR} and \ref{figure:SED-broad} with $\xtejqcq$ in Fig.\,3 of \citeauthor{chaty:2011a} \citeyear{chaty:2011a}). This is consistent with the synchrotron nature for the emission of both systems, which is usually associated with the presence of compact radio jets \citep[see e.\,g.][]{corbel:2002b, chaty:2003b}.
Nevertheless, one main difference between the emission detected from both sources is related to the position at which the spectral slope is broken in the OIR domain. While in $\hdsqt$, a spectral break must be located at longer wavelengths than for the K$_s$ band, the spectral slope of $\xtejqcq$ breaks at higher frequencies at some point between J and I bands. This could be indicative of a more powerful jet present in the latter while a fainter jet would justify the outlier properties of $\hdsqt$, being radio-faint, and therefore also less energetic in NIR.

Regarding the rapid photometry, both $\xtejqcq$ and $\hdsqt$ presented short-term variability in the IR domain in the low, hard state (compare Fig.~\ref{figure:lightc} with Fig.\,5 of \citeauthor{chaty:2011a} \citeyear{chaty:2011a}), especially during the transition state in 2004. 

\section{Conclusion}  \label{section:conclusion}

We conducted OIR photometry, rapid K$_s$ photometry, and IR spectroscopy and built an OIR and multiwavelength SED on the microquasar $\hdsqt$. The observations were taken in HSS states for most of the time, except for the last visit, when the source was in the LHS. We detect fast-time variability in the K$_s$ band, and an emission line in the IR spectra coming from the accretion disk. We discovered a spectral break that must be located at longer wavelengths than the K$_s$ band during the faint 2008 outburst ({\it v}) in the low, hard state, indicating a transition from the optically thick regime towards optical domain to the optically thin regime towards the radio domain. This result is to be added to the short list of microquasars for which a spectral break in the OIR regime has been unambiguously determined. By adding our IR observations of $\hdsqt$ to the OIR vs X-ray correlation of black hole binaries, we showed that the microquasar $\hdsqt$ is radio-quiet and NIR-dim, and that it is the second source, after $\xtejscq$, to be an outlier in the NIR vs X-ray correlation. Finally, we suggested that the companion star is a late-type main sequence star located in the Galactic bulge.

\begin{acknowledgements}

We thank the anonymous referee for a careful reading of the paper and for constructive comments.
SC would like to thank both the ESO staff for performing service observations and Josep Mart\'{\i} Ribas for his hospitality in the Universidad de Jaen, where this work was finalized.
SC is grateful to Guillaume Bacques, Vincent Le Gallo, and Elise Egron for their bibliographic work on $\hdsqt$ during their Masters internship, and to Dave M. Russell for making the data points of the correlation used in Figure \ref{figure:correlation} available.
A.J.M.A. acknowledges support by Consejer\'{\i}a de Econom\'{\i}a, Innovaci\'on, Ciencia y Empleo of Junta de Andaluc\'{\i}a for the research group FQM-322 and excellence fund FQM-5418, as well as FEDER funds.
  {\it Rossi-XTE} Results were provided by the ASM/{\it Rossi-XTE} teams at MIT and at the {\it Rossi-XTE} SOF and the GOF at NASA GSFC. 
  IRAF is distributed by the National Optical Astronomy Observatories, which are operated by the Association of Universities for Research in Astronomy, Inc., under a cooperative agreement with the National Science Foundation.
  This research has made use of NASA's Astrophysics Data System Bibliographic Services and of data products from the Two Micron All Sky Survey, a joint project of the University of Massachusetts and the Infrared Processing and Analysis Center/California Institute of Technology, funded by the National Aeronautics and Space Administration and the National Science Foundation.
  This work was supported by the Centre National d'Etudes Spatiales (CNES), based on observations obtained with MINE --the multiwavelength INTEGRAL NEtwork--, and by the European Community via contract ERC-StG-200911.

\end{acknowledgements}




\end{document}